\documentclass[12pt,preprint]{aastex}

\lefthead{Bekki, Tsujimoto,  \& Chiba}
\righthead{Fates of bulge winds}

\begin{document}

\title{Role of galactic gaseous halos in recycling enriched winds from
bulges to disks:
A new bulge-disk chemical connection}

\author{Kenji Bekki} 
\affil{
School of Physics, University of New South Wales, Sydney 2052, Australia}

\author{Takuji Tsujimoto} 
\affil{
National Astronomical Observatory, Mitaka-shi, Tokyo 181-8588, Japan}

\and

\author{Masashi Chiba}
\affil{
Astronomical Institute, Tohoku University, Sendai, 980-8578, Japan\\}

\begin{abstract}
We  demonstrate for the first time that
gaseous halos of disk galaxies can play a vital role
in recycling  metal-rich gas ejected from the  bulges 
and thus in promoting chemical evolution of  disks.
Our numerical simulations show that metal-rich stellar winds
from bulges in disk galaxies  can be accreted onto the thin disks
owing to  hydrodynamical interaction between the gaseous
ejecta and the gaseous halos, if the mean densities of the halos
(${\rho}_{\rm hg}$)
are as high as $10^{-5}$ cm$^{-3}$. 
The total amount of gas that is ejected from a bulge
through a stellar wind
and then accreted onto the disk
depends mainly on ${\rho}_{\rm hg}$
and the initial velocity of the stellar wind.
About $\sim 1$\% of gaseous ejecta from bulges 
in disk galaxies  of scale length  $a_{\rm d}$ can be accreted onto disks
around $R\sim  2.5 a_{\rm d}$ 
for a reasonable set of  model parameters.
We discuss these results in the context of the origin of the
surprisingly high metallicities of the solar neighborhood disk stars
in the Galaxy.
We also discuss some implications of the present results
in terms of chemical evolution of disk galaxies with possibly
different ${\rho}_{\rm hg}$ in different galaxy environments.
\end{abstract}


\section{Introduction}

Gaseous infall  onto galactic disks from  their halos
has long been considered to play vital roles 
in various aspects of galaxy formation and evolution,
such as formation of S0 galaxies from truncation
of halo gas infall (e.g., Larson et al. 1980),
maintenance of spiral arms in disk galaxies (e.g., Sellwood \& Carlberg 1984),
and  as a solution to the so-called G-dwarf problem (e.g., Pagel 1997).
The total masses and abundance properties
of halo gas infalling onto disks 
and the timescales of gas infall
have been considered  to be key determinants which 
can control galaxy evolution processes (e.g., Sellwood \& Carlberg
1984), in particular, chemical evolution of disk galaxies 
(e.g., Chiappini et al. 1997).
Although it remains largely unclear both observationally
and theoretically how halo gas can be accreted gradually onto disks,
the gradual infall of halo gas
is one of key ingredients in 
most  chemical evolution models
that successfully  explain their chemical  properties
(e.g., Matteucci \&
Fran\c{c}ois 1989).
 
It is well established that  there are
significantly metal-rich disk stars with [Fe/H]$\sim + 0.4$
in the solar neighborhood
(e.g., Feltzing \& Gustafsson 1998; Bensby et al. 2005).
Recent observational and theoretical studies have suggested
that the canonical models  of the Galactic chemical evolution
based on infall of metal-poor halo gas have a serious
problem in explaining the observed presence
of metal-rich stars with [Fe/H] = +0.2$-$0.4 (e.g., Tsujimoto 2007).
One of possible ideas to solve this problem is that the Galactic disk
experienced accretion of metal-rich gas which originated from
the bulge: the observed rather metal-rich disk stars in
the solar neighborhood were  formed from gas ejected from the bulge
(Tsujimoto 2007). 
It is, however, totally unclear whether the gas ejected from the
Galactic bulge
can really reach the solar-neighborhood 
under hydrodynamical influences of the gaseous halo  owing to lack of
extensive numerical studies on dynamical  fates of the ejected bulge gas.

The purpose of this Letter is thus to 
show, for the first time, how gaseous halos of disk galaxies
influence metal-rich stellar winds from the bulges
through nuclear activity (i.e., starbursts and AGN).
We numerically investigate whether and how  stellar winds  
from bulges (or stellar ejecta due to supernova feedback)
can be accreted onto  the disks 
after hydrodynamical interaction between the ejecta
and the gaseous halos.
This investigation is important, not only because
recent observational studies have found evidence of
large-scale outflow from bulges in the Galaxy
and other disk galaxies (e.g.,
Bland-Hawthorn \& Cohen 2003; Matthews \& Grijs 2004;
Veilleux et al. 2005),
but also because the physical role  of galactic gaseous halos
in chemical evolution of disk galaxies has  not been so far
investigated by numerical simulations.
We mainly discuss how the total amount and  distribution of
gas ejected from a bulge  and then accreted onto the 
disk 
depend on physical properties of the gaseous halo
and the bulge ejecta. 
It should be stressed here that the bulge wind would  need to 
have taken place
in the Galaxy several Gyr ago to explain the possible ages of the super
metal-rich solar neighborhood stars  (Tsujimoto 2007).

\section{The model}

We numerically investigate the dynamical evolution of gas ejected from
bulges
in disk galaxies using our own GRAPE-SPH codes as we have done in
previous
studies (e.g., Bekki \& Chiba 2006). A disk galaxy is represented by a
dark
matter halo, gaseous halo, stellar and gaseous disks, a stellar bulge,
and
gas initially within the bulge (referred to as "bulge gas" for
convenience
from now on). The dark matter halo has an "NFW" profile (Navarro et al.
1996) with total mass of $M_{\rm dm}$ ,
 scale length $r_{\rm s}$, and $c$ parameter of 10.
The
gaseous halo has mass $M_{\rm hg}$
 and the same spatial distribution as the dark
matter and is assumed to be initially in hydrostatic equilibrium. The
initial
gaseous temperature of a halo gas particle is therefore determined by
the gas
density, total mass, and gravitational potential at the location of the
particle via Euler's equation for hydrostatic equilibrium (e.g., the
equation
1E-8 in Binney \& Tremaine 1987).

The bulge has a Hernquist profile (Hernquist 1990) with mass $M_{\rm b}$
 and
scale-length $a_{\rm b}$. The disk has an exponential density profile with mass
$M_{\rm d}$,
radius $R_{\rm d}$, and scale length $a_{\rm d}$ (= $0.2R_{\rm d}$).
The stellar and gaseous disks
have
masses of $M_{\rm s}$  and $M_{\rm g}$,
 respectively, and the same exponential density
profile,
and the gas mass fraction ($M_{\rm g} /M_{\rm d}$) is a free parameter. In addition to
the
rotational velocity due to the gravitational field of disk, halo, bulge
components, initial radial and azimuthal velocity dispersion are
assigned to
the disk component according to the epicyclic theory using Toomre's
parameter
(Binney \& Tremaine 1987) Q = 1.5.
We do not include the outflow of stellar winds from stars in the disk
in the present models,
because we think that the stellar winds with lower metallicities
do not significantly influence chemical evolution of the disk
in comparison with the metal-rich bulge wind.

The initial distribution of the bulge gas with
 mass  $M_{\rm bg}$ is assumed 
to be the same as that of the bulge and each bulge gas particle
is assumed to be ejected from the bulge 
in the radial direction
with the ejection velocity  $v_{\rm ej}$.
Also the particles are assumed to have initial rotational
velocities  of $v_{\rm rot}$ (around  the $z$-axis)
ranging from 0 to a circular
velocity at $R=5a_{\rm b}$.
We show the results of the models
with  $v_{\rm rot}=212$ km s$^{-1}$,
because the results do not depend strongly on $v_{\rm rot}$. 
All bulge gas particles are assumed to have the same
initial $v_{\rm ej}$ and $v_{\rm rot}$ in each simulation.
The initial temperature of the bulge gas ($T_{\rm bg}$) is assumed to be
$10^4$ K, which corresponds to warm ionized gas observed in
starburst galaxies such as  M82 (e.g., Heckman et al. 1987).
The models with very high temperatures
of  $T_{\rm bg} \sim 10^6$ K do not show significant gas accretion onto disks.

Since we mainly discuss the abundance properties of the Galactic disk,
we chose the following values of the model parameters
for disk galaxies: 
$M_{\rm dm}=10^{12} {\rm M}_{\odot}$,
$r_{\rm s}/a_{\rm d}=5$,
$M_{\rm d}=6 \times 10^{10} {\rm M}_{\odot}$,
$a_{\rm d}=3.5$ kpc,
$f_{\rm g}=0.1$,
$M_{\rm b}=10^{10} {\rm M}_{\odot}$,
and $a_{\rm b}=0.7$ kpc.
The most important parameters in the present study
are the mean halo gas density
${\rho}_{\rm hg}$ within $3R_{\rm d}$ determined from
$M_{\rm hg}$ and $v_{\rm ej}$ for a given $M_{\rm bg}$.
Although we have investigated many models with different
${\rho}_{\rm hg}$ and $v_{\rm ej}$,
we show the results of the standard model M1
with ${\rho}_{\rm hg}=10^{-5}$ cm$^{-3}$, 
$v_{\rm ej}=500$ km s$^{-1}$,
and $M_{\rm bg}=10^{8} {\rm M}_{\odot}$.
Strong bulge winds can be more clearly seen
 in the models 
with $v_{\rm ej} \ge 500$ km s$^{-1}$.
Sembach et al. (2003) showed that 
the Galaxy gaseous halo is highly extended ($R>70$ kpc)
and low-density (${\rho}_{\rm hg} \le 10^{-4} - 10^{-5}$ cm$^{-3}$),
which means that the adopted  ${\rho}_{\rm hg}$ is reasonable.
The total number of particles used in a simulation ranges from
228805 to 258805 depending on $f_{\rm g}$ and $M_{\rm hg}$.

Parameter values of
the  models (M1 to M7) discussed in this paper are shown in Table 1.
We do not  discuss how rotation of gaseous halos
influences the accretion processes of bulge gas onto disks,
because we find that the influence is not significant.
We show some results of comparative models in which
$f_{\rm g}=0$ and ${\rho}_{\rm hg}=0$ cm$^{-3}$: these models
enable us to grasp essential roles of gaseous halos in the accretion
processes,
though they are unrealistic. 
We will describe in detail the results of models not discussed 
in this
paper (e.g., those with different 
$M_{\rm bg}$ and $v_{\rm rot}$) in our future papers.
We mainly investigate the final total mass of bulge gas particles
settled around  ``the solar neighborhood'' ($R=R_{\odot}\sim2.5a_{\rm d}$) 
with 7 kpc $\le R \le$ 10 kpc and $|z| \le 1$ kpc ($M_{\rm acc}$)
for each model.
We also investigate the mass ratio ($f_{\rm acc}$) of $M_{\rm acc}$ to
$M_{\rm bg}$ for each model.
The final gaseous distributions at $T=0.45$ Gyr correspond to those
of the Galaxy about several Gyr ago in the present study.

\section{Results}

Fig. 1 shows how gas ejected from a bulge evolves with time
during hydrodynamical interaction between the gas
and the halo gas in a disk galaxy for the standard model M1.
Although a significant  fraction of the gas once escapes
from the bulge ($T=0.11$ Gyr),  most of the gas can be 
finally returned 
to the disk plane
owing to gaseous pressure from the halo and the disk
($T=0.45$ Gyr).
The bulge gas particles initially in the central region of the bulge
can interact so strongly  with the disk gas particles immediately
after the ejection  that  a thin disk can be formed 
from the bulge gas at $R<3$ kpc  for a short
time scale ($T=0.06$ Gyr).
The bulge gas particles ejected from the bulge regions
with larger $|z|$ 
can be later accreted onto the outer part of the disk
and consequently can  rotate around the center of the galaxy. 
The mean orbital eccentricity of bulge gas particles in
the solar neighborhood is 0.08 in this model, which means that
the orbits become almost circular, because
the particles have acquired  orbital angular momentum during
hydrodynamical interaction with the disk gas. 

Fig. 2 shows that  $M_{\rm acc}$ increases with time
owing to rapid accretion of the bulge gas 
and the mean accretion rate within  0.45 Gyr
is  $2.2 \times 10^{-3} {\rm M}_{\odot}$ yr$^{-1}$
in the standard model.
The mass-ratio ($f_{\rm acc}$) of  $M_{\rm acc}$ to  $M_{\rm bg}$
is $\sim 0.01$, which means that a small fraction of
the bulge gas can be accreted onto the solar neighborhood.
Fig. 2 also shows that this result of  small $M_{\rm acc}$ 
is true for the comparative model M2 with reasonable ${\rho}_{\rm hg}$,
which suggests that the bulge gas can not so efficiently  be accreted onto
the disk outside the solar neighborhood irrespective of ${\rho}_{\rm
hg}$. The more rapid accretion and smaller $M_{\rm acc}$
in the model M2 is due largely to the stronger hydrodynamical
interaction between the bulge  and the halo gas.
Fig. 3 shows that almost 80\% of the bulge gas can be accreted
onto the (inner) disk region with $R\le 3$ kpc in the standard model: 
this {\it preferential
accretion} can be clearly seen also in the model M2.

Fig. 4 compares the final distributions of bulge gas particles
at $T=0.45$ Gyr between different four models (M$2-5$). 
Only a  compact thin disk with $M_{\rm acc}=1.4 \times 10^4 {\rm
M}_{\odot}$
can be developed from the bulge gas in the model M3 without
halo gas, which clearly demonstrates that hydrodynamical interaction
between  bulge and halo gas in a disk galaxy
is crucial for the accretion
of the bulge gas onto the outer part of the disk. 
Model  M4 contain no  disk and halo gas 
(i.e., $f_{\rm g}=0$ and ${\rho}_{\rm hg}=0$ cm$^{-3}$)
and  does not show a thin disk,
which means that the presence
of gas disks is also important for recycling of the bulge
ejecta in disk galaxies.
This result of M4 combined with that of M1 confirms 
that hydrodynamical interaction between
bulge and disk gas is essential for  forming a {\it rotating}  gas disk
composed of the bulge gas.

The bulge
gas can be pushed back to the inner disk more strongly
and quickly by the halo gas owing to stronger hydrodynamical
pressure of the gaseous halo  against the bulge gas in the model M2.
In this case  the bulge gas can be  accreted only onto the inner disk
rather than the outer one
in M2: only a very small fraction of the bulge gas can reach
the solar neighborhood 
and
outer gaseous streams and small gas clumps seen in the standard model
are not  seen in M2.
The small gas clumps seen in the standard model
might well be identified as high-velocity clouds,
if they can form H~{\sc i}  gas owing to efficient cooling.
Less energetic gaseous flow from the bulge
can be pushed back to the disk by the 
gaseous halo  so that 
a thin disk with no
small gas clumps 
is  formed 
in the model M5.

The dependences of $f_{\rm acc}$ on model parameters 
are summarized as follows.
Firstly,   there is an optimum $v_{\rm ej}$ which
maximizes $f_{\rm acc}$ for a reasonable 
${\rho}_{\rm hg}=10^{-5}$ cm$^{-3}$.
A larger fraction of bulge gas can escape from the disk
for larger $v_{\rm ej}$ so that a smaller fraction
of the ejecta is finally accreted onto the solar neighborhood
in the model M6 with $v_{\rm ej}=1000$ km s$^{-1}$.
On the other hand, 
a smaller fraction of the bulge gas can reach $R=R_{\odot}$ 
owing to less energetic flow so that a smaller fraction
of the gas can be accreted onto the solar neighborhood
in  model M5 with $v_{\rm ej}=200$ km
s$^{-1}$. 
Thus,  $f_{\rm acc}$ is  maximum
for moderately energetic stellar winds such as
$v_{\rm ej}=500$ km s$^{-1}$.

Nevertheless,   $f_{\rm acc}$ can be higher for higher ${\rho}_{\rm hg}$
in the models with $v_{\rm ej}=1000$ km s$^{-1}$:
in this case, higher densities  of the halo gas are  required for preventing the 
bulge gas from escaping the disk for 
more energetic stellar winds. This ${\rho}_{\rm
hg}$-dependence of $f_{\rm acc}$ is totally different from
that seen for  $v_{\rm ej}=500$ km s$^{-1}$,
which implies that $f_{\rm acc}$ depends on ${\rho}_{\rm hg}$
and  $v_{\rm ej}$ in a complicated way.
The model M4 with no halo/disk gas shows 
$f_{\rm acc}$ (0.18) higher than that in model M3 with disk gas.
Such high $f_{\rm acc}$ in M4 is due to bulge gas
particles that happen
to be around the solar-neighborhood but  are not rotating the disk:
these particles soon return back to their original locations.
It should be finally noted  that the present results do not depend strongly on
$v_{\rm rot}$.

\section{Discussion and conclusions}

Our results indicate for the first time 
that  gaseous halos of disk galaxies
are very  important for chemical evolution of the disks
in the sense that  they enable the galaxies to
recycle the metal-rich stellar winds
from the bulges and accelerate the chemical enrichment of galactic  disks.
The derived  
preferential
accretion of the gaseous ejecta onto the inner disks will lead to a  
steepening
of the abundance gradient after the action of strong central  
starbursts in the bulges followed by
stellar winds. This steep abundance gradient is predicted to flatten  
with time
owing to a subsequent chemical enrichment under an accretion of 
low-metallicity
infall from the halo (Tsujimoto et al. 2008), which is in accord with  
the observed flattening
feature in radial metallicity gradients
over the last several Gyr (Chen et al. 2003; Maciel et al.  
2006, 2007).

Our important finding is that the metal-rich gaseous ejecta from the  
bulge can
reach and chemically enrich the solar neighborhood through a one-time 
sporadic
accretion event triggered by a starburst. This mechanism offers
a promising  
channel to produce
super-metal-rich stars with $+0.2<$[Fe/H]$<+0.4$, the presence of  
which can not be
accounted for  by the conventional scheme of Galactic chemical evolution  
models (Tsujimoto 2007). 
An important caveat is that
the observed very metal-rich stars with [Fe/H]$ \sim +0.4$ would need to
be formed directly from the bulge gas without significant
dilution with metal-poor halo and disk gas.

Using the observed metallicity distribution function  by
Nordstr\"om et al. (2004), we  estimate that
the mass fraction of metal-rich stars with $+0.2<$[Fe/H]$<+0.4$
is about 4\% in the solar annulus.
This means that the possible total {\it stellar} mass ($M_{\rm mr}$)
of the metal-rich stars 
for $8$ kpc $\le R \le$
$9$ kpc (i.e., the solar annulus)
is about $3 \times 10^7 {\rm M}_{\odot}$
for the adopted Galactic disk mass of $6 \times 10^{10} {\rm M}_{\odot}$
and the exponential scale
length of 3.5 kpc. 
The total  mass of the metal-rich stars 
thus would give some constraints on (i) the total mass of the initial
bulge wind (thus the strength of the central starburst)
and (ii) to what degree the metal-rich bulge gas (e.g.,
[Fe/H]$\sim$+0.4) needs to be diluted via mixing of halo/disk
gas to form less metal-rich stars with [Fe/H]$\sim$+0.2.

Our study shows 
that about 1 \% of gas ejected from bulges in disk galaxies can be 
accreted onto the disks at $R\sim 2.5 a_{\rm d}$ for 
a reasonable set of model parameters in the present study. 
This  suggests that chemical evolution
of the outer parts ($R\sim 3a_{\rm d}$) of the disks can not be 
so strongly influenced  by the accretion of the metal-rich gaseous ejecta.
This also implies that if infalling metal-rich gas with [Fe/H]$\sim$+0.4
in the Galaxy  can be mixed
well with the gas with [Fe/H] $\sim 0$ and
a similar mass already present
in the solar neighborhood and then form new stars
from the mixed gas,
the new stars would have [Fe/H]$\sim$+0.2: they however 
are  unlikely to
exhibit super-solar metallicities with [Fe/H]$\sim +0.4$. 
However, if new stars can form preferentially from the metal-rich
gas {\it during the accretion of the gas onto the disk}
for some physical reasons, 
formation of new stars with super-solar metallicities would be possible.
Our simulations imply that 
(i) dilution of the metal-rich bulge gas is likely
to occur after the accretion of the gas onto the Galaxy
and (ii) the dilution due to interaction with halo gas
would be  unlikely owing to rapid accretion of the bulge gas.

Previous chemical evolution models of disk galaxies did not consider
{\it rapid accretion of metal-rich gas ejected from bulges onto
disks} (e.g., Lacey \& Fall 1985; 
Prantzos \& Aubert 1995; 
Chiappini et al. 1997). 
The present study suggests that future more sophisticated 
chemical evolution models need to  
include possible influences of the rapid accretion
on  chemical evolution
for disk galaxies, 
in particular,  for those with bigger bulges which would have experienced
stronger nuclear starbursts.
The present study has not clarified (i) local (pc-scale)
star formation processes during the  accretion of metal-rich gas
onto disks and (ii) metallicities of new stars formed from the star
formation.  We plan to investigate numerically 
how mixing processes of metal-poor
disk gas and metal-rich bulge ejecta within disks determine the 
metallicities of new stars formed during the accretion of the ejecta.

\acknowledgments
We are  grateful to the anonymous referee for valuable comments,
which contribute to improve the present paper.
K.B. acknowledges the financial support of the Australian Research
Council throughout the course of this work.
The numerical simulations reported here were carried out on GRAPE
systems kindly made available by the Center for Computational
Astrophysics (CfCA)
at National Astronomical Observatory of Japan (NAOJ).

\newpage

\begin{deluxetable}{ccccc}
\footnotesize
\tablecaption{Model parameters and brief results.}
\tablewidth{0pt}
\tablehead{
\colhead{  model no} &
\colhead{  ${\rho}_{\rm hg}$ (cm$^{-3}$) } &
\colhead{  $v_{\rm ej}$ (km s$^{-1}$) } &  
\colhead{  $f_{\rm acc}$  ($\times 10^{-2}$) } &  
\colhead{  comments} }
\startdata
M1 & $10^{-5}$ &  500  & 1.03  & the standard model   \\
M2 & $10^{-4}$ &  500  & 0.39  &    \\
M3 & $0$ &  500  & 0.02  & without halo gas  \\
M4 & $0$ &  500  & 0.18  & without halo/disk gas ($f_{\rm g}=0$)   \\
M5 & $10^{-5}$ &  200  & 0.22  &    \\
M6 & $10^{-5}$ &  1000  & 0.18  &    \\
M7 & $10^{-4}$ &  1000  & 0.55  &    \\
\enddata
\end{deluxetable}

\clearpage

\begin{figure}
\epsscale{0.8}
\plotone{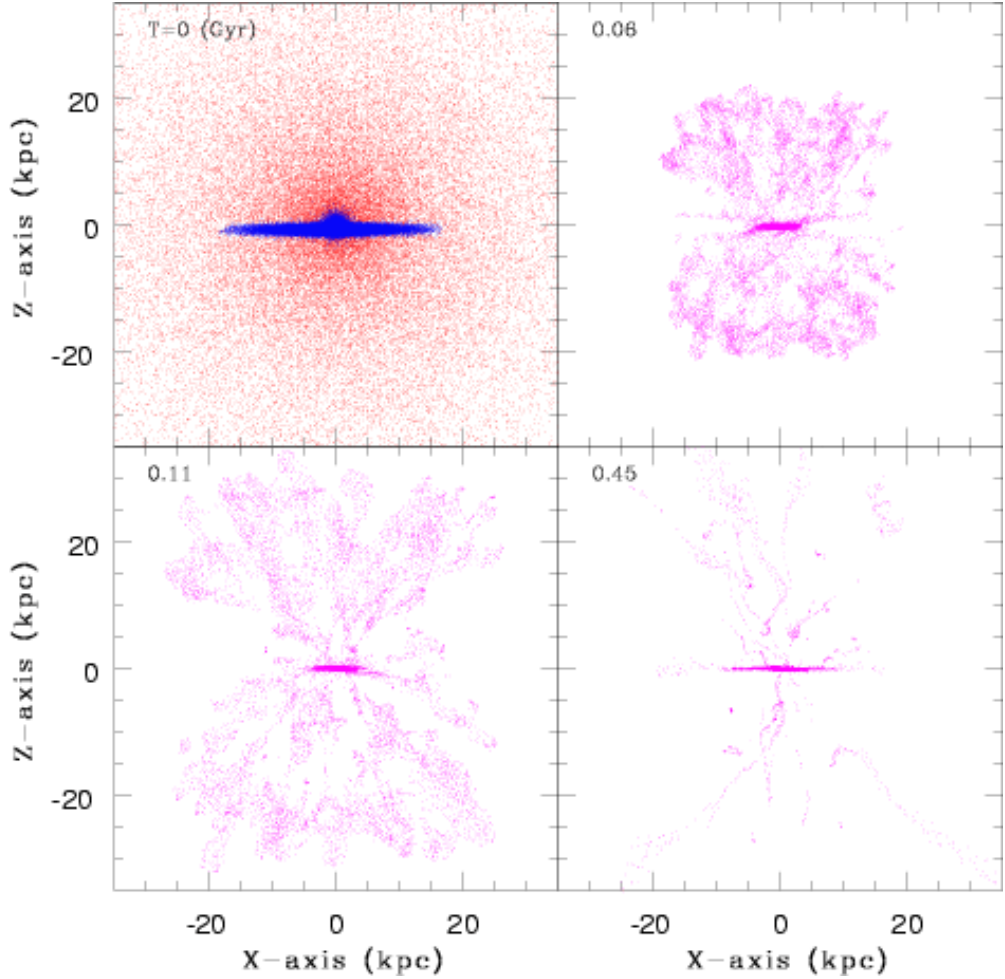}
\figcaption{
Time evolution of the
mass distribution in a disk galaxy projected onto the $x$-$z$ plane
for the standard model M1 with ${\rho}_{\rm hg}=10^{-5}$ cm$^{-3}$.
The time $T$ in units of Gyr is given in the upper left corner
for each panel. The mass distributions of disk stellar  particles
and halo  gaseous ones
are shown by blue and red, respectively, in the upper
left panel for $T=0$.
For other panels,  only  the distribution of gas ejected from the bulge
(i.e., ``bulge gas particles'')
is shown by magenta so that the time evolution of the distribution can be
more clearly seen.
Note that most of the bulge gas  particles  
can  finally settle down to the thin disk after hydrodynamical
interaction between the bulge and halo gas. 
\label{fig-1}}
\end{figure}

\begin{figure}
\epsscale{0.8}
\plotone{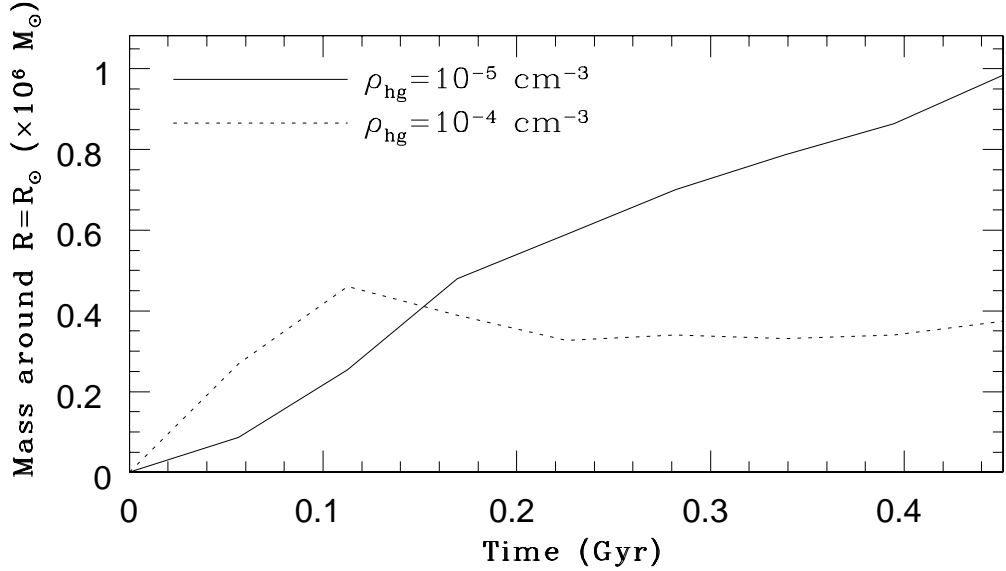}
\figcaption{
The time evolution of the total amount of bulge gas particles
accreted onto the thin disk with 7 kpc $\le R\le 10$ kpc
(i.e., around the solar neighborhood, $R=R_{\odot}$)
and with $|z|\le 1$ kpc
in the standard model 
with ${\rho}_{\rm hg}=10^{-5}$   cm$^{-3}$ (solid) and the model M2 with  
 ${\rho}_{\rm hg}=10^{-4}$   cm$^{-3}$ (dotted).
\label{fig-2}}
\end{figure}

\begin{figure}
\epsscale{0.8}
\plotone{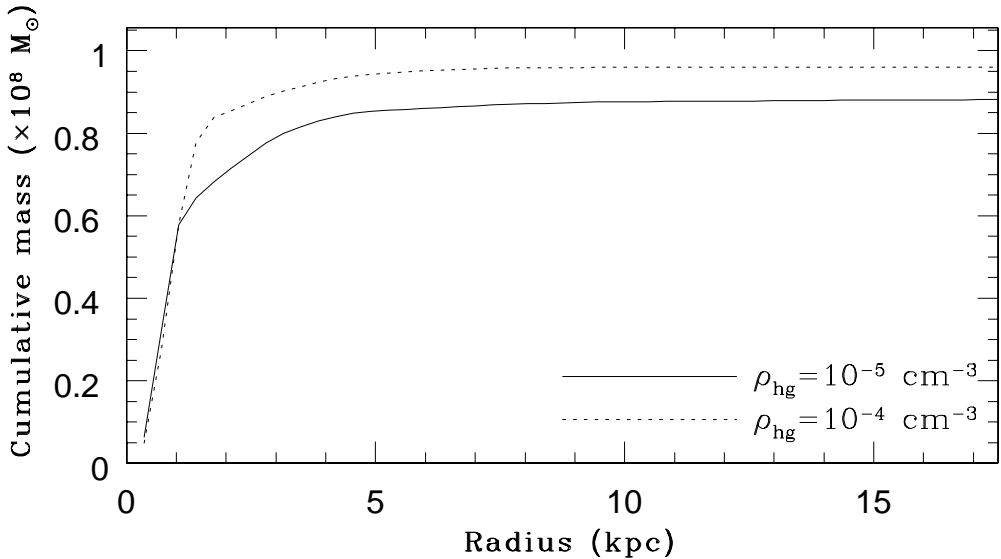}
\figcaption{
The cumulative mass of bulge gas particles within $R$
that are located within the thin disk
(i.e., $|z|\le 1$ kpc)  at $T=0.45$ Gyr for the standard model
(solid) and the model M2 (dotted).
\label{fig-3}}
\end{figure}

\begin{figure}
\epsscale{0.8}
\plotone{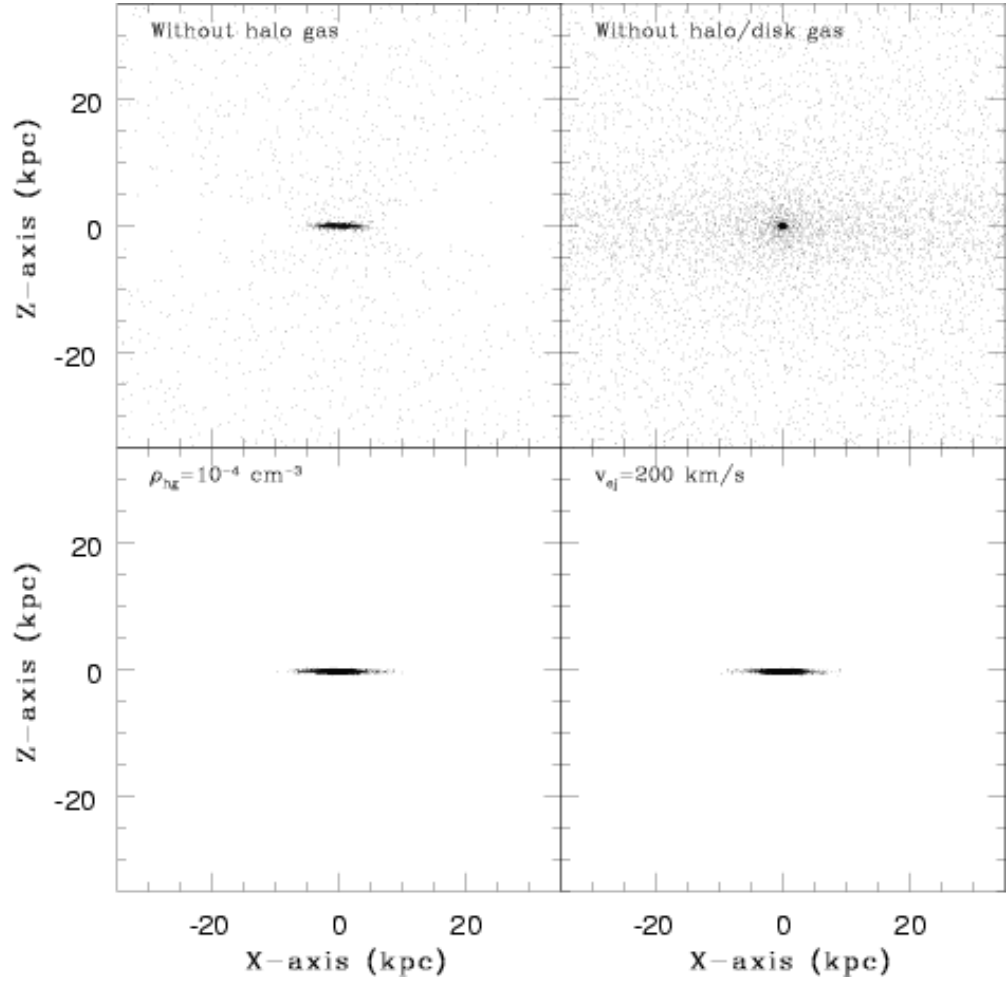}
\figcaption{
Final distributions of bulge gas particles projected onto the $x$-$z$
plane at $T=0.45$ Gyr for the model M3 without halo gas (upper left),
M4 without halo/disk gas (upper right),
M2 with ${\rho}_{\rm hg}=10^{-4}$   cm$^{-3}$ (lower left), 
and M5 with $v_{\rm ej}=200$ km s$^{-2}$ (lower right).
\label{fig-4}}
\end{figure}
\end{document}